# 2D pn junctions driven out-of-equilibrium†

Ferney A. Chaves, [ORCID]* Pedro C. Feijoo [ORCID] and David Jiménez [ORCID]

The pn junction is a fundamental electrical component in modern electronics and optoelectronics. Currently, there is a great deal of interest in the two-dimensional (2D) pn junction. Although many experiments have demonstrated the working principle, there is a lack of fundamental understanding of its basic properties and expected performances, in particular when the device is driven out-of-equilibrium. To fill the current gap in understanding, we investigate the electrostatics and electronic transport of 2D lateral pn junctions. To do so we implement a physics-based simulator that self-consistently solves the 2D Poisson's equation coupled to the drift-diffusion and continuity equations. Notably, the simulator takes into account the strong influence of the out-of-plane electric field through the surrounding dielectric, capturing the weak screening of charge carriers. Supported by simulations, we propose a Shockley-like equation for the ideal current–voltage ($J$–$V$) characteristics, in full analogy to the bulk junction after defining an effective depletion layer (EDL). We also discuss the impact of recombination–generation processes inside the EDL, which actually produce a significant deviation with respect to the ideal behavior, consistently with experimental data. Moreover, we analyze the capacitances and conductance of the 2D lateral pn junction. Based on its equivalent circuit we investigate its cut-off frequency targeting RF applications. To gain deeper insight into the role played by material dimensionality, we benchmark the performances of single-layer $MoS_2$ (2D) lateral pn junctions against those of the Si (3D) junction. Finally, a practical discussion on the short length 2D junction case together with the expected impact of interface states has been provided. Given the available list of 2D materials, this work opens the door to a wider exploration of material-dependent performances.

## 1. Introduction

The need for miniaturization of structures in modern semiconductor electronics has directed significant attention toward the fabrication of pn junctions based on atomically thin materials such as graphene, phosphorene, silicene, germanene, *etc.*, and the family of transition metal dichalcogenides (TMDs).[1–7] This kind of 2D pn junction takes advantage of the ultra-thin nature of 2D materials to offer new and exciting possibilities which are impossible to achieve by its 3D counterpart. These 2D pn junctions will be an essential part of the new generation of 2D crystal based electronic and optoelectronic devices such as photodiodes, transistors, solar cells, photo-detectors, *etc.*[8–23] Some advantages of 2D pn junctions are the thickness-dependency of the semiconductor band gap for some materials, the gate-tunable doping and rectification by external out-of-plane electric fields and the possibility of tuning some electrical properties such as the depletion region width and capacitance by changing the surrounding dielectrics. These advantages enable even more possibilities to create different pn junction concepts, compared with 3D materials.

2D pn junctions can be integrated into a device either by using an out-of-plane configuration, featuring vertically stacked van der Waals-bonded layers,[5,14,21,24] or by a lateral (in-plane) configuration, spanning adjacent regions of different doping in the covalently bonded 2D plane.[15–18,23,25–27] At the same time, they can be classified as homojunctions (based on a single 2D material) and heterojunctions (formed by joining two different 2D materials). The 2D homojunctions can be (1) thickness-based junctions,[22] (2) elemental doping based junctions,[5,28] (3) electrostatically doped junctions,[11,15,29,30] and (4) chemically doped junctions, in which there is surface chemical doping or substitutional doping of a region in a 2D material.[16,31–35]

In particular, 2D lateral pn junctions (2DJ) are useful for solar cells and photodetectors because the built-in potential created in the extended space charge region separates and drives the photogenerated e–h pairs to generate photocurrent.[25–27]

While there are many reports of experimental studies on 2D junctions, the same is not true for theoretical studies, where the impact of low dimensionality on their electrical properties and their differences with 3D junctions (3DJs) are reported. A few theoretical studies have contributed to the quantitative understanding of the electrostatics in the equilibrium of 2DJs.[36–45]

*Departament d'Enginyeria Electrònica, Escola d'Enginyeria, Universitat Autònoma de Barcelona, Campus UAB, 08193 Bellaterra, Spain. E-mail: ferneyalveiro.chaves@uab.cat*

Most of them have been focused on chemically doped symmetric pn junctions, achieving analytical expressions for the electrostatic potential profile and the depletion width with dependency on physical parameters such as doping densities and the dielectric constant of the surrounding media. However, due to the presence of a very extended transition region between the fully depleted layer and the quasi-neutral region in 2DJs, in comparison with 3DJs, inheriting a weak screening of charge carriers, it is truly complex to obtain a closed analytical theory in the out-of-equilibrium regime, as exists in 3DJs. Thus, theoretical research studies of 2D pn junctions driven out-of-equilibrium, considering the impact of its low dimensionality, remain to be developed.

Here, we report a study of the 2D pn junction driven out-of-equilibrium based on numerical simulations, which sets the basis for gaining understanding of experimental measurements and for the assessment of future analytical models. Our study has been focused, specifically, on the chemically doped 2D lateral pn junction (2DJ), leaving the investigation of electrostatically doped and vertical 2D pn junctions for a future study. The manuscript has been divided in the following sections: in Section 2, we have described the numerical model used to investigate the properties of the 2DJ out-of-equilibrium; and in Section 3 we have proposed a simple model describing the ideal electrical characteristics of the 2DJ. Deviations from the ideality have also been discussed. In Section 4 we have investigated the impact of dimensionality on the $J$–$V$ characteristics, specifically we have benchmarked the monolayer MoS$_2$ junction against the silicon junction. In Section 5, we have addressed the calculations of capacitances and conductance of the 2DJ out-of-equilibrium. Based on these results, we have investigated its FoM, such as the cut-off frequency. In Section 6 the effect of the interface states on the electrical performances is addressed and finally in Section 7 we have discussed the role played by the dielectric environment in defining the $J$–$V$ characteristics and the possibility of tuning the diode rectification.

## 2. Model

Fig. 1 represents the physical structure of the 2DJ considered in this work. The 2DJ, made of an abrupt junction of p-type and n-type semiconductors at $x = 0$, lies in the plane $z = 0$. The device along the $y$-axis is large enough to consider the independence of its properties in that direction. The metal electrodes at the edges of the device make perfect ohmic contacts and supply a ground reference potential to the right edge and a potential $V$ to the left edge.

A 2D non-linear Poisson's equation describes the electrostatics of the device, involving the electrostatic potential distribution in the $xz$-plane $\phi(x,z;V)$, the space charge density $\sigma = q(p - n + N_D - N_A)$ and the effective dielectric permittivity $\varepsilon_{eff}$ of the insulating media surrounding the junction. Strictly, out-of-equilibrium electrostatics is coupled with the 1D charge transport along the lateral dimension $x$ by means of the drift-diffusion and continuity equations. The transport involves non-linear first-order differential equations for the electron (hole) quasi-Fermi potential $V_{n(p)}(x)$ and current density $J_{n(p)}$, and contains physical parameters such as the electron (hole) mobility $\mu_{n(p)}$, the minority electron (hole) lifetime $\tau_{n(p)}$, the energy gap of the semiconductor $E_g$ and the temperature $T$.[46] The in-plane potential $\varphi(x) = \phi(x,0;V)$ and the quasi-Fermi potentials $V_{n(p)}$ are related to the local intrinsic energy $E_i$ and electron (hole) Fermi energy as $E_i(x) = -q\varphi$ and $E_{Fn(p)}(x) = -qV_{n(p)}$, respectively. Processes of generation and recombination of charge carriers, mediated by capture and emission of carriers at trap centers located in the band gap of the semiconductor, are modeled with a net recombination rate $U$ obeying the Shockley–Read–Hall model. A complete description of the coupled equations modeling the physics of the 2DJ driven out-of-equilibrium and details about the algorithm to numerically solve them are given in the ESI.† All the simulations in this work have been carried out assuming room temperature to be $T = 300$ K.

Fig. 2 shows the band diagram, quasi-Fermi energies and intrinsic energy profiles of a 2DJ, as calculated from our numerical model. In equilibrium a built-in potential $\varphi_{bi}$ develops across the depletion layer, while in the out-of-equilibrium regime it is reduced/augmented as $\varphi_{bi}-V$ according to the sign of $V$.

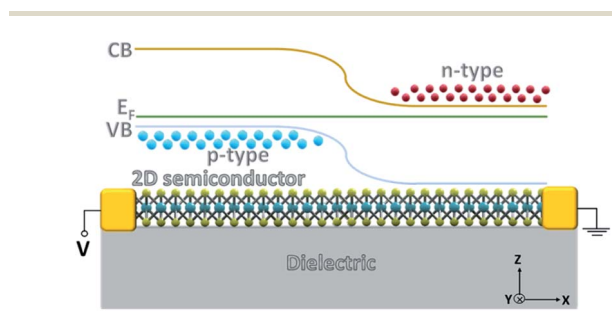

**Fig. 1** Scheme of the 2D lateral pn junction geometry.

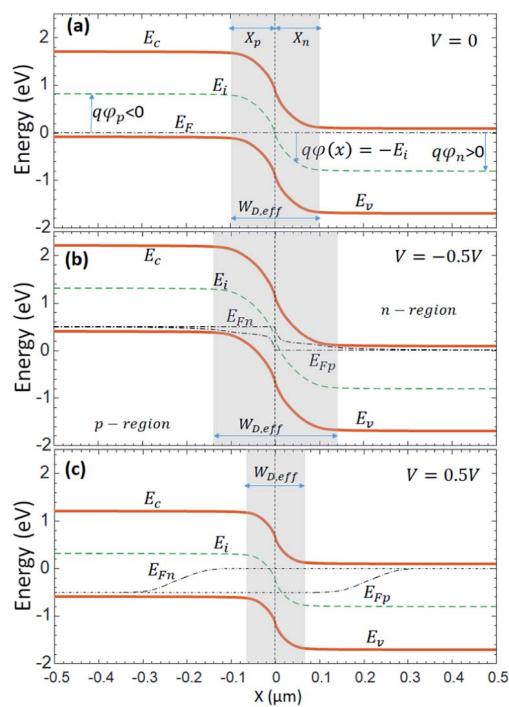

**Fig. 2** Band diagram of a 2D lateral pn junction in thermal equilibrium (a); reverse biased (b); and forward biased (c).

## 3. The ideal 2D diode

It is possible to obtain an analytical model (equivalent to Shockley's equation) for the ideal 2DJ, like in conventional 3DJs. For 3DJs the following approximations are made to develop the ideal model:[47,48] (i) a dipole layer fully depleted of carriers with abrupt boundaries supports the voltage drop $\varphi_{bi}-V$, outside those boundaries, the semiconductor is assumed to be neutral. (ii) The carrier statistics is assumed to follow a Boltzmann distribution, implying that the semiconductor is assumed to be non-degenerate. (iii) The excess of minority carriers injected in the quasi-neutral region is much less than the corresponding majority carrier density (low injection approximation). (iv) There is a zero net recombination rate ($U = 0$) inside the depletion layer such that the total current is only due to the diffusion component of minority carriers in the quasi-neutral regions. In this section, we discuss the possibility of extending the previous assumptions to 2DJs in order to obtain the 2D equivalent to the Shockley equation. This will be done on the basis of numerical simulations relying on the model described in the ESI.†

Fig. 3 shows, from our simulations, the normalized charge density along the n-side of symmetric 2DJ and 3DJ devices in thermal equilibrium. These devices have similar parameters which correspond to 2DJ1 and 3DJ1 described in Table 1, except for their corresponding DOS. Remarkably, the semiconductor of the 2DJ1 device has an energy gap similar to that of Si and equivalent doping characterized by the energies $q\varphi_n$ and $q\varphi_p$ represented in Fig. 2a, i.e. $\varphi_{n,2D} = \varphi_{n,3D} = |\varphi_{p,2D}| = |\varphi_{p,3D}|$. In Fig. 3, the shaded regions represent the transition regions between the fully depleted layer ($\sigma/qN_D \sim 1$) and the quasi-neutral region ($\sigma/qN_D \sim 0$) for both devices. It is worth noting that we have used the notation $\sigma$ and $N_D$ for both the 2D case and 3D case, that is, $\sigma$ and $N_D$ represent surface (volume) charge density and surface (volume) doping density in the 2D (3D) case, respectively. The length of the transition regions is $L_{2D} \sim 48$ μm and $L_{3D} \sim 0.1$ μm for the examined 2DJ and 3DJ, respectively. In the ESI† the possible impact of the edge effects for several device lengths is reported, concluding it to be low. The 2D transition region is much wider ($\sim 500\times$ for the examined case) than the 3D transition region because of a weaker screening of charge carriers related to a significant out-of-plane electric field, as compared with the strong screening exhibited in 3DJs. The strong screening in 3DJs justifies the total depletion approximation, giving rise to a well-defined depletion region of width:

$$W_D = \sqrt{\frac{4\varepsilon(\varphi_{bi} - V)}{qN}}, \qquad (1)$$

where $N = N_A = N_D$ for the symmetric junction. For 2DJs it is still possible to define an effective depletion layer (EDL) width analogous to the 3D case, which allows us to obtain an equivalent Shockley model. One of the most relevant theoretical studies reporting an analytical expression for the EDL in both symmetric and asymmetric 2DJs has been carried out by Nipane et al., which uses a method based on an infinite number of reflections of the image charge.[38] The expression for the given EDL is

$$W_{D,eff(1)} = \frac{\pi^2 \varepsilon_{eff}(\varphi_{bi} - V)}{-qN_D f(N_A/N_D)} \qquad (2a)$$

where

$$f(\xi) = \sum_{k=1}^{\infty} \frac{\left[1 - (-1)^k\right](1 + \xi)\sin\left(\frac{\pi k}{1 + \xi}\right)}{k^2} \qquad (2b)$$

The half of the values given by eqn (2a) corresponds to $\sigma/qN = 56.4\%$ for the symmetric case (see Fig. 3). However, eqn (2) is inaccurate for asymmetric 2DJs when compared with our numerical simulations. We have found that the numerically determined depletion width $W_{D,\text{eff}(2)}$ under the condition $\sigma/qN = 56.4\%$ is more correct for the general case,[36] guaranteeing an almost zero field at the edges of the EDL (see the ESI† for a detailed comparison between $W_{D,\text{eff}(1)}$ and $W_{D,\text{eff}(2)}$).

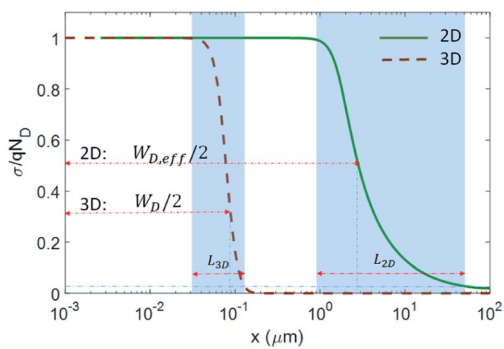

**Fig. 3** Charge density profile for symmetric 2D lateral and 3D pn junctions, showing the depletion region and transition region for both cases. Only the n-side region is shown.

Table 1 Parameters used in our simulations. SL: single layer

| Device | Description | $E_g$ (eV) | $\varepsilon_{eff}/\varepsilon_0$ | $n_i$ (cm$^{-2}$) | $\mu_p$ (cm$^2$ V$^{-1}$ s$^{-1}$) | $\mu_n$ (cm$^2$ V$^{-1}$ s$^{-1}$) | $\tau_p$ (s) | $\tau_n$ (s) | $\phi_p$ (eV) | $\phi_n$ (eV) |
|---|---|---|---|---|---|---|---|---|---|---|
| 2DJ1 | Sym. pn. Small gap | 1.12 | 11.9 | $5 \times 10^3$ | 280 | 350 | $1 \times 10^{-6}$ | $2 \times 10^{-7}$ | 0.4 | 0.4 |
| 2DJ2 | Asym. p$^+$n. Large gap (SL MoS$_2$) | 1.8 | 3.9 | $9.9 \times 10^{-3}$ | 50 | 50 | $3 \times 10^{-10}$ | $3 \times 10^{-10}$ | 0.89 | 0.81 |
| 2DJ3 | Sym. pn. Large gap (SL MoS$_2$) | 1.8 | 3.9 | $9.9 \times 10^{-3}$ | 50 | 50 | $3 \times 10^{-10}$ | $3 \times 10^{-10}$ | 0.81 | 0.81 |
| 3DJ1 | Sym. pn. Low doping (Si) | 1.12 | 11.9 | $1.4 \times 10^{10}$ | 280 | 350 | $1 \times 10^{-6}$ | $2 \times 10^{-7}$ | 0.4 | 0.4 |
| 3DJ2 | Sym. pn. High doping (Si) | 1.12 | 11.9 | $1.4 \times 10^{10}$ | 280 | 350 | $1 \times 10^{-6}$ | $2 \times 10^{-7}$ | 0.47 | 0.47 |
| 3DJ3 | Asym. p$^+$n (Si) | 1.12 | 11.9 | $1.4 \times 10^{10}$ | 350 | 350 | $2 \times 10^{-7}$ | $2 \times 10^{-7}$ | 0.55 | 0.47 |

Once the appropriate EDL is defined, it is possible to consider the 2DJ as consisting of three regions as in 3DJs: the quasi-neutral p-side, the quasi-neutral n-side, and the EDL. Now, by assuming zero net recombination ($U = 0$) inside the EDL, low injection, and the Boltzmann approximation to calculate the carrier densities, a Shockley-like equation for the ideal 2DJ case can be obtained. Although not shown here, the 2D Shockley-like equation is obtained, in total analogy with the 3D case,[4] from the solution of the continuity equation to obtain the diffusion current of the minority carriers $J_{n(p)}$ in the quasi-neutral regions, which allows the determination of the total current ($J_{\text{ideal}}$) density for a given bias as:

$$J_{\text{ideal}} = J_p + J_n = J_o[e^{qV/kT} - 1], \tag{3}$$

where

$$J_o = qn_i^2 \left[ \frac{D_p}{L_p N_D} + \frac{D_n}{L_n N_A} \right] \tag{4}$$

is the reverse saturation current density per unit length. Here, $n_i \cong g_{2D}kT \exp(-E_g/2kT)$ is the intrinsic carrier density, with $g_{2D}$ the band-edge density of states (DOS). The electron (hole) diffusion coefficient and diffusion length are defined as $D_{n(p)} = kT\mu_{n(p)}/q$ and $L_{n(p)} = \sqrt{D_{n(p)}\tau_{n(p)}}$, respectively. For 2D semiconductor crystals $n_i$ is typically much smaller than $10^{11}$ cm$^{-2}$ of zero-gap graphene. For example, monolayer MoS$_2$ and MoTe$_2$ semiconductors, with gaps of 1.8 eV and 1.15 eV, respectively, have intrinsic carrier densities $\sim 10^{-2}$ cm$^{-2}$ and $\sim 10^{3}$ cm$^{-2}$, respectively.[46]

Next, the ideal 2D diode equation has been benchmarked against numerical simulations assuming $U = 0$ inside the EDL. Fig. 4a presents the results for the symmetric case. The parameters used here correspond to the device 2DJ1 described in Table 1, for which $L_n \sim 27$ μm and $L_p \sim 13.5$ μm. As can be observed from the figure, the 2D Shockley-like equation (ideality factor $\eta = 1$) fits very well with the numerical simulation at both reverse and forward polarizations even when the full depletion approximation is not strictly applicable. On the other hand, Fig. 4b shows the ideal $J$–$V$ characteristics together with the numerical simulations for an asymmetric 2D p$^+$n junction with parameters described for device 2DJ2 in Table 1. Again, the black dashed curve has been calculated with the Shockley-like equation and the others are from the numerical model with the two different definitions given for $W_{D,\text{eff}}$. Importantly, the 2D Shockley-like equation fits the numerical simulation much better taking $W_{D,\text{eff}(2)}$ as the effective depletion layer width rather than $W_{D,\text{eff}(1)}$.

Next, we have analyzed in more detail the electrostatics of the asymmetric 2DJ. Fig. 5a shows the relative total charge density calculated at forward bias as a function of the position. The asymmetric 3DJ case is also shown for comparison purposes. The parameters used here are those of the 2DJ2 and 3DJ3 devices in Table 1. Clearly, the widths $W_{D,\text{eff}(1)}$ and $W_{D,\text{eff}(2)}$ are quite different. The reason why $W_{D,\text{eff}(2)}$ results in a better fit of the Shockley model with the numerical model can be understood by means of Fig. 5b, which shows the electrostatic potential along the device. At $W_{D,\text{eff}(2)}$ the electric field is much closer to zero compared to $W_{D,\text{eff}(1)}$, so the beginning of the quasi-neutral region is better defined with $W_{D,\text{eff}(2)}$. Fig. 5c shows that the electric field for the 2D case is not linear with position, as in the 3D case, but it is described by $\ln(x/(x - X_n))$ near the junction,[38] which comes from the contribution of the lines of charge in which the block of sheet charge within the space charge region can be split, each of those lines contributing an electric field $\propto 1/x$. This is in striking contrast with the 3D case where the contribution of the vertical planes of charge produces a linear electric field inside the depletion region. Finally, Fig. 5d shows the carrier density profiles $n(x)$ and $p(x)$ of the 2DJ2 device at equilibrium and out-of-equilibrium. In the inset, we have repeated the carrier density profiles at forward bias ($V = 0.5$ V) in a wider region, exhibiting a constant slope (in logarithmic scale) in the quasi-neutral region, near the edge of the EDL, related to the hole diffusion length $L_p$. The extracted value of $L_p$ is $\sim 0.19$ μm which coincides with $L_p \equiv \sqrt{kT\mu_p\tau_p/q}$, as expected from the parameters of the 2DJ2 device in Table 1.

## 4. Impact of the dimensionality in 2D lateral pn junctions

In order to investigate the impact of the dimensionality in 2DJs, we have carried out simulations of both symmetric 2DJ and 3DJ with similar parameters as given by devices 2DJ1 and 3DJ1, respectively, exceptfor their DOS. We have considered R–G processes inside the EDL. The devices have been simulated up to forward biases such that the series resistance of the quasi-neutral regions can be observed. The resulting current densities are shown in Fig. 6, normalized to the corresponding $J_o$. In the main panel, marked as "Ideal", are the $J$–$V$ characteristics with $U = 0$ inside the EDL defined by $W_D$ and $W_{D,\text{eff}}$ in eqn (1) and (2), respectively. The "Ideal" curves fit the Shockley equation $J = J_o(e^{qV/\eta kT} - 1)$, with an ideality factor $\eta = 1$, quite well within the voltage range where no appreciable influence of series resistance is observed. More specifically, as shown in the inset, the ideality factor extracted from the numerical results is $\sim 1$ up to $V \sim 0.7$ V for the "Ideal" case, and then it quickly increases due to the resistance of the quasi-neutral regions. For the calculations we have not taken into account any breakdown mechanism such as tunneling, thermal instability or avalanche multiplication because of the low electric fields at the considered low reverse bias.[48,49] On the other hand, the "Non-ideal" curves are obtained after considering $U \neq 0$ inside the EDL. Different regions of $J$–$V$ characteristics can be observed depending on the bias, namely: generation current ($G$) at reverse bias, recombination current ($R$) at low forward bias, diffusion current ($D$) at moderate forward bias, and current dominated by the series resistance ($R_s$) at high forward bias. High injection conditions are not present within the considered voltage range. Clearly, some differences between the 2D and 3D cases appear, even though the used parameters are similar. In the main figure, at reverse bias, the relative current density of the 2DJ is larger than that of the 3DJ. To understand this result, let us consider, firstly, the current driven by R–G processes under reverse-bias. Inside the EDL $p \ll n_i$, $n \ll n_i$ so $U < 0$ is then dominated by generation of electron–hole pairs, and can be expressed as:

$$U \approx -\frac{n_i^2}{n_i \left\{ \tau_p \left(1 + \frac{n}{n_i}\right) + \tau_n \left(1 + \frac{p}{n_i}\right) \right\}} \cong -\frac{n_i}{\tau_g}, \quad (5)$$

where $\tau_g$ is the generation lifetime, defined by the expression in brackets. The current due to generation in the effective depletion region can be roughly estimated by assuming a mean value $\overline{U} = -n_i/\overline{\tau_g}$ of net recombination inside the EDL

$$J_G = \int_{-X_p}^{X_n} q|\overline{U}|dx \approx \frac{q n_i W_{D,\text{eff}}}{\overline{\tau_g}} \quad (6)$$

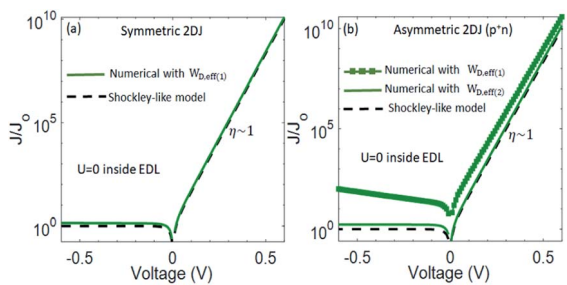

**Fig. 4** Ideal $J$–$V$ characteristics of the 2D lateral pn junction. (a) Symmetric 2DJ1 device. (b) Asymmetric 2DJ2 device. $W_{D,\text{eff}(1)}$ has been calculated according to eqn (2) and $W_{D,\text{eff}(2)}$ according to the criterion $\sigma/qN \sim 0.56$. In both cases a zero net recombination rate ($U = 0$) has been assumed inside the EDL.

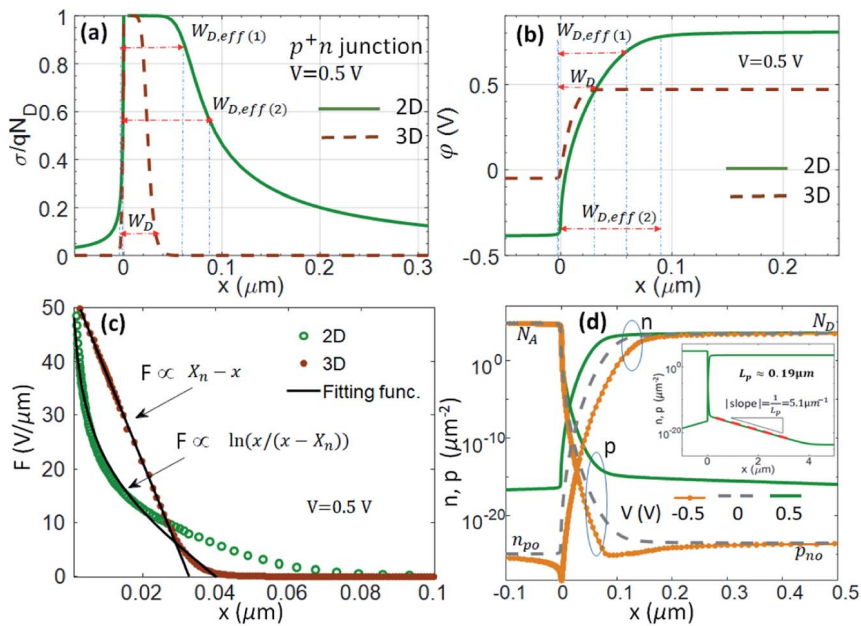

**Fig. 5** Electrostatic properties of the asymmetric 2D lateral pn junction. (a) Normalized total charge; (b) electrostatic potential; (c) electric field; (d) carrier density (p and n) in the 2DJ at equilibrium, and forward and reverse bias. Inset: Numerical extraction of the hole diffusion length. The asymmetric 3DJ case is also shown in (a)–(c) for comparison purposes.

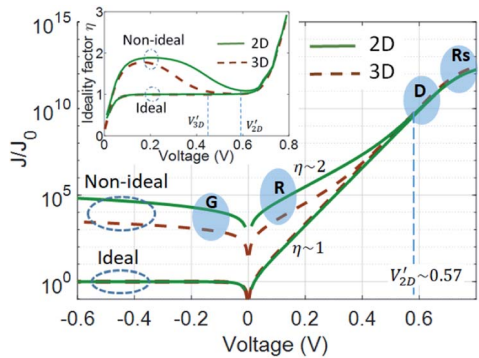

**Fig. 6** $J$–$V$ characteristics for symmetric 2D lateral and 3D pn junctions showing the different components of the current. Inset: Ideality factor vs. V. The labels "$G$", "$R$", "$D$" and "$R_s$" correspond to the regions dominated by the generation current, recombination current, diffusion current and current driven by series resistance, respectively.

Fig. 7 shows the profiles of $U$ for a 2DJ at both reverse and forward bias, obtained from our numerical model, along with $|\overline{U}|$ in the EDL with $\overline{\tau_g} \sim \tau_p \sim \tau_n$. Thus, the ratio $J_G/J_o$ can be written as

$$\frac{J_G}{J_E} \propto \frac{N_D}{n_i} \frac{W_{D,\text{eff}}}{(L_p + L_n)}. \quad (7)$$

This result is also valid for 3DJs after exchanging $W_{D,\text{eff}}$ with $W_D$. Typically, $W_{D,\text{eff}}$ for 2DJs is larger than $W_D$ in 3DJs because of the weaker screening of charge carriers in 2DJs. On the other hand, $L_p$ and $L_n$ are typically smaller in 2DJs as a consequence of smaller lifetimes and mobilities. Assuming the same $N_D/n_i$ for the 2DJ and 3DJ, it does result in a larger $J_G/J_o$ for the 2D case.

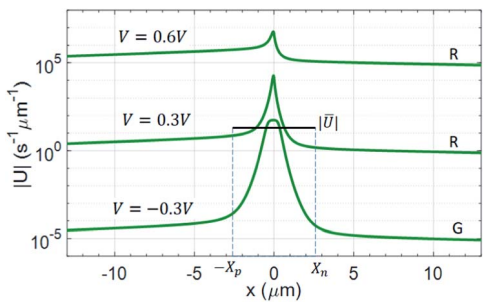

**Fig. 7** Net recombination rate profile at reverse bias ($V = -0.3$ V) and forward bias ($V = 0.3$ V and 0.6 V).

On the other hand, at forward bias, the major R–G processes in the EDL are the electron and hole capture processes (recombination) from the conduction and valence bands, respectively, at the trap centers in the bandgap. Thus, a net recombination rate $U > 0$ produces a dominant recombination current ($R$) over the diffusion current ($D$) in the quasi-neutral regions. This is valid from zero bias up to the crossover voltage $V'_{2D}$ reached when $J_R/J_{ideal} \sim 1$ (see Fig. 6). By following the same procedure in the literature for 3D pn junctions, it is possible to find a simple expression for the recombination current density $J_R$ at $V > kT/q$:[50]

$$J_R = \sqrt{\frac{\pi}{2}} \frac{kTn_i}{\tau F_o} \exp\left(\frac{qV}{2kT}\right),\qquad(8)$$

where $F_o$ is the electric field at the location of maximum recombination and there is a characteristic ideality factor $\eta = 2$. To get this expression for $J_R$ we have assumed $\tau_p = \tau_n = \tau$ for simplicity. As shown in Fig. 5c, the dependence of the electric field for the 2D on the position is not as simple as in the 3D case and, in general, numerical approximations are needed to find $F_o$ as a function of the applied voltage. However, by using the simple approximation $F_o \approx \bar{F} = \dfrac{\phi_{bi} - V}{W_{D,eff}}$, which is the average electric field in the EDL, a bias independent formula $F_o = \dfrac{4GgqN_D}{\pi^2 \varepsilon_{eff}}$, valid for symmetric junctions, where $G = 0.915$ is Catalan's constant, can be obtained. Thus, the ratio $J_R/J_{ideal}$ at forward bias results in

$$\frac{J_R}{J_{ideal}} \approx \sqrt{\frac{\pi}{2}} \frac{kT}{q} \frac{\pi^2 \varepsilon_{eff}}{8Gqn_iL_p} \exp\left(-\frac{qV}{2kT}\right).\qquad(9)$$

Again, this ratio is typically larger for 2DJs than for 3DJs due to the lower intrinsic carrier densities, carrier lifetimes and mobilities for 2D semiconductors. In addition, a closed expression for the crossover voltage $V'_{2D}$, can be obtained:

$$V'_{2D} \approx \frac{2kT}{q} \log\left(\sqrt{\frac{\pi}{2}} \frac{kT}{q} \frac{\pi^2 \varepsilon_{eff}}{8Gqn_iL_p}\right).\qquad(10)$$

Using the parameters for the device 2DJ1, $V'_{2D} \sim 0.57$ V agrees with the obtained value in Fig. 6 from our numerical model. The value of $V'_{3D}$ cannot be analytically obtained, but can be obtained from our simulation $V'_{3D} \sim 0.42$ V, as illustrated in the inset in Fig. 6. These effects can also be understood through the difference between their EDL in thermal equilibrium with $W_{D,eff} = 5.24$ μm and $W_D = 0.17$ μm for the 2DJ and 3DJ, respectively.

Given the larger transition region $L_{2D}$ exhibited by the 2DJ compared with the 3DJ, as shown in Fig. 3, it might be possible that the physical length of the 2DJ is shorter than the diffusion length. This special case, which is of practical interest, has been discussed in the ESI.†

Now, let's examine the differences between a typical 3DJ made of Si and a 2DJ made of monolayer MoS₂. For this, we have simulated symmetric devices 2DJ3 and 3DJ2 described in Table 1. For the sake of making a fair comparison the selected doping in each device should fulfil $(E_F - E_V)|_{2D} = (E_F - E_V)|_{3D} = (E_c - E_F)|_{2D} = (E_c - E_F)|_{3D}$ in the quasi-neutral region, far away from the EDL. The numerical results of the normalized $J$–$V$ characteristics are shown in Fig. 8. We can observe that $J_G/J_o$ for the 2DJ is ~10⁸ larger than the 3D one, and the crossover voltage $V'_{2D}$ for the 2D pn junction is approx. 1 V larger than $V'_{3D}$. Again, this can be understood because of the drastic reduction of both the intrinsic carrier density and diffusion length of minority carriers (smaller lifetimes and mobilities) together with the larger EDL in the 2D case. As for the 2D case the effect of $R_s$ is not visible within the considered voltage range. However, for the 3DJ the $R_s$ effect starts to be strong at ~0.8 V. Remarkably, the MoS₂ 2DJ exhibits an ideality factor close to 2 up to ~1.3 V, indicating the dominance of the recombination current over the diffusion current. Interestingly, experimental evidence of this phenomenon has been reported by Choi et al. for 7 nm thick chemically doped MoS₂ lateral pn junctions and Baugher et al. for electrically tunable pn diodes.[15,16] Their experimental evidence reveals that the thinner the crystal semiconductors the nearer to 2 the ideality factor.

## 5. Capacitances and the cut-off frequency in 2D lateral pn junctions

In this section we address the investigation of the capacitances associated with the 2DJ in the out-of-equilibrium regime. An equivalent circuit will be presented and the cut-off frequency of the 2DJ will be analyzed. For the sake of simplicity, we have focused the discussion on a symmetric 2DJ. Let us begin discussing the depletion-layer capacitance per unit length, defined as $C_{dp} = dQ_D/dV$, where $dQ_D$ is the incremental depletion charge on each side of the junction after applying an incremental voltage change $dV$. By assuming the total depletion approximation, the depletion charge on the p-side, for example, can be calculated using $Q_D = -qNW_{D,eff}(V)/2$. From eqn (2a) we can obtain:

$$C_{dp} = \frac{\pi^2 \varepsilon_{eff}}{8G}. \tag{11}$$

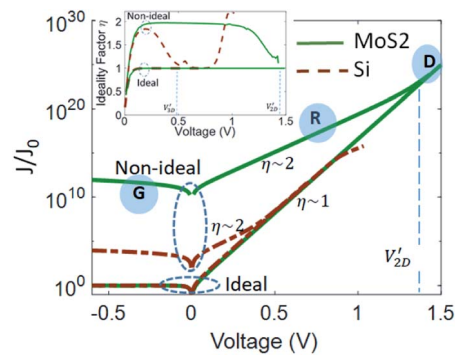

**Fig. 8** $J$–$V$ characteristics of symmetric pn junctions: MoS₂ junction (2D case) and Si junction (3D case). Inset: Ideality factor vs. voltage.

$C_{dp}$ does not depend on the applied voltage, in striking contrast to the 3D case. The depletion-layer capacitance accounts for most of the junction capacitance when the junction is reverse-biased. At forward bias, a diffusion capacitance exists in addition to the depletion capacitance. The diffusion capacitance is the result of the rearrangement of minority surface carriers in the quasi-neutral regions. Thus, in a similar way to the 3D case, when a small ac signal, characterized by an angular frequency $\omega$, is applied to a 2DJ that is forward-biased at a dc voltage $V$, the total voltage and current can be described as

$$V^*(t) = V + v^* \exp(i\omega t) \tag{12}$$

$$J^*(t) = J + j^* \exp(i\omega t) \tag{13}$$

where $v^*$ and $j^*$ are the small-signal voltage and current density, respectively. The real and imaginary parts of the admittance $j^*/v^*$ are related to the diffusion conductance $G_d$ and diffusion capacitance $C_{df}$, respectively. By applying the same procedure as the one used for 3DJs,[48] the following expressions for the low frequency diffusion capacitance and conductance, respectively, are obtained:

$$C_{df} = \frac{q^2 n_i^2 (L_p + L_n)}{2kTN} \exp\left(\frac{qV}{kT}\right), \tag{14}$$

and

$$G_d = \frac{q^2 n_i^2}{kTN}\left(\frac{D_p}{L_p} + \frac{D_n}{L_n}\right)\exp\left(\frac{qV}{kT}\right), \tag{15}$$

which are the same results as the one that can be obtained by differentiating eqn (3), that is $G_d = dJ_{ideal}/dV = r_d^{-1}$, where $r_d$ is defined as the differential resistance. In Fig. 9 we have shown the behavior of both $C_{dp}$ and $C_{df}$ capacitances as a function of $V$ for 2DJ1 with a small gap (Fig. 9a) and 2DJ3 with a large gap (Fig. 9b). More details of these devices are given in Table 1. The green solid lines correspond to the capacitances obtained from our numerical model, according to

$$C_{dp} = \frac{d}{dV}\int_0^{W_{D,eff}/2} q\sigma(x)dx, \tag{16}$$

and

$$C_{df} = \frac{d}{dV}\int_{W_{D,eff}/2}^{x_{max}} q\Delta p(x)dx + \frac{d}{dV}\int_{x_{min}}^{-W_{D,eff}/2} q\Delta n(x)dx, \tag{17}$$

where $\Delta p(x)$ and $\Delta n(x)$ are the excess of hole and electron surface densities, respectively, in the quasi-neutral regions. The black solid lines refer to the theoretical model, where the $C_{dp}$ and $C_{df}$ capacitances have been calculated from eqn (11) and (14), respectively. In Fig. 9 we have included the corresponding 3DJ simulations and theoretical model outcome for comparison purposes (brown dashed lines).[48] The 2DJ and 3DJ capacitances have units of $F/\mu m$ and $F/\mu m^2$, respectively. Importantly, eqn (11) and (14) fit very well to the numerical results, both in 2D and 3D cases.

By taking advantage of the $C_{dp}$ insensitivity to the voltage, a simple formula for the critical bias $V^*$ that makes $C_{dp} \cong C_{df}$, can be obtained:

$$V* = \frac{kT}{q}\log\left(\frac{kT}{q^2}\frac{\pi^2 \varepsilon_{eff}}{4Gn_i^2}\frac{N}{(L_p + L_n)}\right) \tag{18}$$

For the examined devices 2DJ1 and 2DJ3, the calculated critical voltages are ∼0.66 V and ∼1.5 V, respectively. These values agree with the numerical results shown in Fig. 9. In general, 2DJs made of large gap semiconductors with both low carrier lifetimes and mobilities exhibit a relatively high critical voltage $V^*$, resulting in a total capacitance $C_T$ dominated by the depletion component, that is $C_T = C_{dp} + C_{df} \sim C_{dp}$ that, interestingly, could be modulated by the dielectric permittivity of the surrounding media, according to eqn (11).

The equivalent circuit model of the diode can be obtained by the total capacitance $C_T$, in parallel with its ac differential resistance $r_d$. The resistance of the quasi-neutral regions and contact resistance can be modeled by a resistance $R_s$ in series with the parallel combination of $C_T$-$r_d$ as shown in the inset in Fig. 10b. It is worth noting that the units of $R_s$ and $r_d$ are $\Omega\ \mu m$ ($\Omega\ \mu m^2$) for 2D (3D) junctions, respectively. Once the diode equivalent circuit is determined, the cut-off frequency can be defined as the frequency at which $v^*/j^* = r_d$

$$f_c = \frac{1}{2\pi C_T(r_d \| R_s)} \tag{19}$$

Fig. 10a shows $f_c$ as a function of the dc voltage for both 2DJ and 3DJ ignoring the extrinsic contact resistance. For the 2DJ case, we have considered the devices 2DJ1 (small gap) and 2DJ3 (large gap), which have been compared with equivalent 3D devices 3DJ1 and 3DJ2, respectively, satisfying $(E_F - E_V)_{2D} = (E_F - E_V)_{3D} = (E_c - E_F)_{2D} = (E_c - E_F)_{3D}$. At low bias ($V < 0.4\ V$), $f_c$ exhibits the highest values for both 2DJ and 3DJ cases. However in this range of voltages $r_d \gg R_s$ and $C_{dp} \gg C_{df}$, and the cut-off frequency can be approximated as $f_c \approx 1/2\pi C_{dp}R_s$. Our simulation shows that the 2DJ1 device, which has the same gap as the 3DJ1 device, exhibits $f_c$ almost two orders of magnitude less than that of its 3D counterpart at low bias. By choosing large gap semiconductors like the monolayer $MoS_2$ the maximum $f_c$ can reach the GHz range, although far away from the frequencies reached by 3DJs, as exemplified by the 3DJ2 device.

As shown in Fig. 10b we have analyzed the influence of doping in defining $f_c$. Specifically, it shows the zero bias cut-off frequency for 2DJ3 and 3DJ2 devices as a function of doping density which has been characterized by $(E_c - E_F)/kT$. Given that we have considered symmetrical junctions, then $(E_F - E_V)|_{p\text{-side}} = (E_c - E_F)|_{n\text{-side}}$. The results obtained reveal the beneficial effect of high doping to get the highest possible $f_c$. According to our numerical result, the silicon based 3DJ can reach zero bias cut-off frequencies over tens of GHz; meanwhile the MoS$_2$ based 2DJ can reach ~1 GHz. Further investigation will be needed with calibrated parameters to determine the potential of 2DJs.

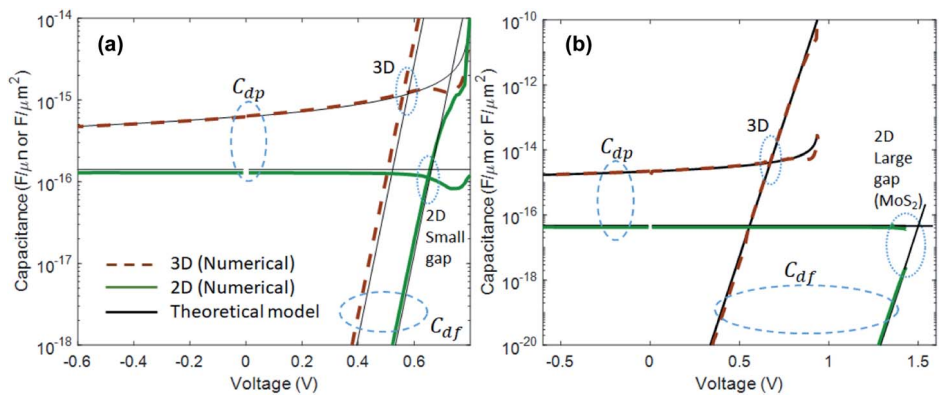

**Fig. 9** Depletion-layer and diffusion capacitances of 2D lateral pn junctions and comparison with the 3D case. (a) Small gap case, represented by devices 2DJ1 and 3DJ1. (b) Large gap case, represented by devices 2DJ3 and 3DJ2.

## 6. Impact of interface states

Understanding the interface properties between the atomically thin semiconductors and dielectrics is fundamentally important for enhancing the carrier transport properties. The impact of interface states on the electrical properties of 2D material based MOSFETs has been addressed in some studies.[51–53] However, as far as we know, the dependence of properties such as diode rectification, capacitance or cutoff frequencies on the density of interface states ($D_{it}$) has not been reported yet and deserves further investigation. Here, we have included a simple model for the interface states between the 2D semiconductor and its surrounding dielectric media, in order to shed light on their effects. We add an interface charge $\sigma_{int}$ to the space charge density $\sigma$ when solving the Poisson's equation

$$\sigma_{int}(\varphi, V_n, V_p) = -q^2 D_{it}\left(\varphi - \frac{V_n + V_p}{2}\right) + \sigma_{int,0} \quad (20)$$

where $D_{it}$ is the density of interface defects (eV$^{-1}$ cm$^{-2}$) and $\sigma_{int,0}$ is the constant sheet density at the interface (C cm$^{-2}$). The details of the model and additional simulations are described in the ESI.†

Fig. 11 shows the impact of the interface states on the current and capacitance characteristics for a symmetric 2DJ with parameters of 2DJ1, from our augmented numerical model. Here we have modified $D_{it}$ from 0 up to $3 \times 10^{10}$ eV$^{-1}$ cm$^{-2}$ and kept $\sigma_{int,0} = 0$ in order to get symmetric 2DJs in every case. In the figure, the arrows indicate the direction of increasing $D_{it}$, which produces a wider EDL (see the inset in Fig. 11a). However, as a rectifier diode we have predicted a slight reduction of the on/off ratio with the increasing of $D_{it}$ (main panel of Fig. 11a). Our numerical results predict no significant effects on the capacitances within the range of considered $D_{it}$, leaving invariant the cutoff frequency. In the ESI (Section S4†) we have included numerical results for asymmetric p$^+$n junctions.

## 7. Impact of the surrounding dielectric permittivity

In this section, we have analyzed the impact of the surrounding dielectric medium on the $J$–$V$ characteristics of 2DJs. Because of

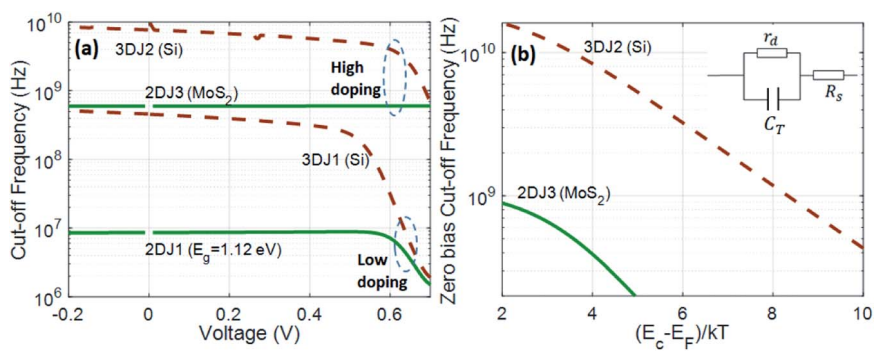

**Fig. 10** (a) Cut-off frequency as a function of the dc bias for 2DJ and 3DJ devices. (b) Zero bias cut-off frequency as a function of the doping for 2DJ (MoS$_2$) and 3DJ (Si) devices, represented by 2DJ3 and 3DJ2, respectively.

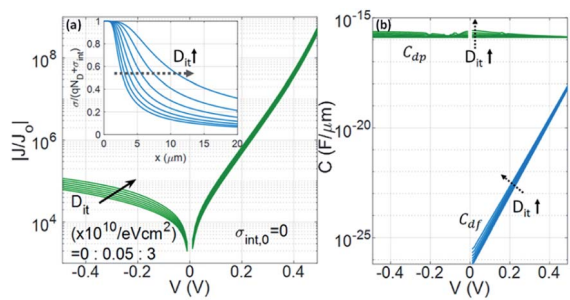

**Fig. 11** Impact of the interface states on electrical characteristics of the 2DJ1 device. (a) $J$–$V$ characteristics normalized to the reverse saturation current given by eqn (4). Inset: Normalized total charge. (b) Depletion and diffusion capacitances. The arrows indicate the direction of increasing $D_{it}$.

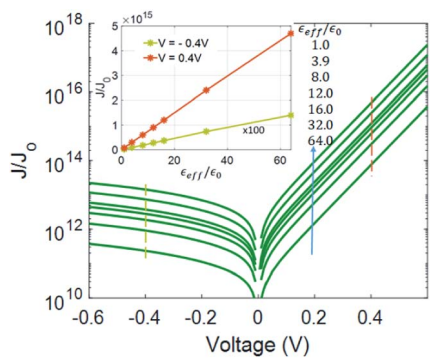

**Fig. 12** Impact of the surrounding dielectric permittivity $\varepsilon_{eff}$ on the $J$–$V$ characteristics of 2DJs.

the bi-dimensionality 2DJs are quite sensitive to the surrounding dielectric, so it is possible to design devices with tailored depletion capacitance (eqn (14)) and current. Eqn (6) and (8) show how the current driven by R–G processes inside the EDL depends on $W_{D,eff}$ and the maximum lateral electric field $F_0$, which are both dependent on the dielectric constant of the surrounding media. Fig. 12 shows the dependence of the relative current density on the effective dielectric constant $\varepsilon_{eff}$ for the symmetric device 2DJ3 described in Table 1. We have assumed that the dielectric–semiconductor interface is free of charge density. The inset in Fig. 12 shows the dependence of the relative current on $\varepsilon_{eff}$ at both reverse bias and forward bias, exhibiting a linear relationship as predicted using eqn (6) and (8), respectively. The results of the effects of interface states on the linear dependence $J/J_0$ vs. $\varepsilon_{eff}$ at both reverse bias and forward bias have been added in the Fig. S6 of the ESI.†

## 8. Conclusions

We have investigated the electrostatics and transport properties of 2D lateral pn junctions driven out-of-equilibrium. For such a purpose we have implemented a physics-based simulator where the 2D Poisson's equation is solved self-consistently with the drift-diffusion and continuity equations. Our numerical simulations reveal the impact of R–G processes inside the depletion layer on the $J$–$V$ characteristics. Deviations from ideal $J$–$V$ characteristics and the impact of the dimensionality have been thoroughly discussed. We have also discussed the possibility of tuning the $J$–$V$ characteristics by using the permittivity of the surrounding environment. Moreover, we have analyzed the capacitances and conductance of 2DJs, which allows us to define an equivalent circuit. Based on this, we have benchmarked the cut-off frequency of 2DJs for different gaps and doping scenarios, taking the 3DJ case as the reference. Finally, a practical discussion on the short length 2D junction case together with the expected impact of interface states has also been provided. The present work opens the door to a wider exploration of potential advantages that 2DJs could bring in terms of FoMs given the long list of available 2D materials. For this to happen a set of material-dependent calibrated parameters is required for any given technology.

## Conflicts of interest

There are no conflicts to declare.

## Acknowledgements

This project has received funding from the European Union's Horizon 2020 Research and Innovation Programme under grant agreements No. GrapheneCore2 785219 and No. GrapheneCore3 881603, and from Ministerio de Ciencia, Innovación y Universidades under grant agreement RTI 2018-097876-B-C21 (MCIU/AEI/FEDER, UE). This article has been partially funded by the European Regional Development Funds (ERDF) allocated to the program Operatiu FEDER de Catalunya 2014–2020, with the support of the Secretaria d'Universitats I Recerca of the Departament d'Empresa I Coneixment of the Generalitat de Catalunya for the Emerging Technology cluster to carry out valorization and transfer of research results. The reference of the GraphCAT project is 001-P-0011702.